# Right Model, Right Task: Capability-Constrained Carbon Labeling to Decarbonize Generative AI

Jiawei Qu[1,2,3], Kai Hou[3*], Hongjie Jia[3], Jinyue Yan[1,2*]

[1] Department of Building Environment and Energy Engineering, The Hong Kong Polytechnic University, Hung Hom, Kowloon, Hong Kong, SAR, China

[2] International Centre of Urban Energy Nexus, The Hong Kong Polytechnic University, Hong Kong, SAR, China

[3] State Key Laboratory of Smart Power Distribution Equipment and System, Tianjin University, Tianjin, China

[*] Corresponding author at: j-jerry.yan@polyu.edu.hk (Jinyue Yan).

## Abstract

The carbon footprint of generative-AI inference depends not only on hardware efficiency and electricity supply, but also on whether the selected model is appropriate for the task. A frontier model may expend unnecessary computation on a routine request, whereas an under-capable model may increase energy use through failure, retries, extended reasoning and escalation. However, a standardized evaluation framework that accounts for task-specific capability requirements and the carbon emissions associated with successful task remains lacking. In this Perspective, we argue for a capability-constrained carbon labeling framework that shifts environmental assessment from carbon intensity per inference towards the expected carbon emissions per successfully completed task at the required capability level. Under this framework, models would first be assessed against task-specific capability requirements, and only models capable of meeting those requirements would be compared on their expected energy use and operational carbon emissions. By linking service quality to carbon cost, this approach could enable more informed matching of the right model to the right task, offering a demand-side pathway to reducing emissions associated with generative-AI token use. Realizing this vision will require standardized capability-based service assessments, comprehensive energy accounting and transparent carbon disclosures.

## Introduction

Generative-AI decarbonization has concentrated on cleaner electricity and more efficient hardware, software and data centres [1]-[3]. However, these measures overlook a demand-side inefficiency: model-task mismatch. This mismatch takes two forms: over-capability and under-capability (Fig.1a). First, an over-capability model may waste energy on a routine task if a less resource-intensive model can meet the same requirement. Similarly, an under-capable model can also be inefficient if failures trigger retries, longer reasoning or escalation. Unsuccessful Qwen3-Coder-30B-A3B runs on SWE-bench Verified consumed more than twice as much energy on average as successful runs [4]. Therefore, the relevant question is not which model is smallest or most capable, but which model completes a specified task successfully with the lowest carbon emissions—the right model for the right task.

Existing model-based routing studies have demonstrated, at the algorithmic level, the feasibility of matching tasks with models of different capabilities [5]. However, the ability to route tasks does not, by itself, establish which model can deliver the required quality of service at the lowest carbon cost. Such comparisons require two elements that current evaluation practices do not consistently provide: task-specific capability thresholds that establish service equivalence in generative AI, and comprehensive energy accounting that captures the resources required for successful task completion.

First, although existing benchmarks of generative AI cover diverse domains, including coding and writing, Existing benchmark scores do not establish the performance thresholds defining different capability levels within a task category. For example, a model may perform well on coding benchmarks yet consistently fail to generate code that passes all required tests for a particular task [6]. Task-specific capability levels should therefore be defined by explicit performance thresholds and success criteria, enabling the identification of models that can reliably meet the required capability level. Second, energy or carbon emissions per token or per query can underestimate the resources required to deliver a successful service, because task failures may induce retries, repeated inference or escalation to more capable models, each adding further energy use and emissions [4]. Therefore, energy consumption metrics should be evaluated per successfully completed task at a defined capability level.

In this Perspective, we argue for a capability-constrained carbon labeling framework that shifts model evaluation from carbon intensity towards the carbon emission of successfully completing a task at the required capability level, thereby supporting the principle of matching the right model to the right task (Fig. 1b). Such a framework should treat task capability as a constraint rather than a quantity to be traded directly against emissions: models that cannot reliably satisfy the requirements of a task should not be considered lower-carbon alternatives simply because they consume less energy per query. Among models that meet the required capability level, comparison should instead reflect the expected energy and carbon emissions associated with successful task completion, including the additional burden of failures and retries. Carbon labels could further incorporate the electricity carbon intensity of serving regions, enabling carbon-aware routing where latency, privacy and availability constraints permit. Rather than assigning fixed rankings to rapidly evolving models, such labels would provide a common basis for comparing models at explicitly defined capability levels and for directing each task towards the lowest-carbon option capable of completing it reliably.

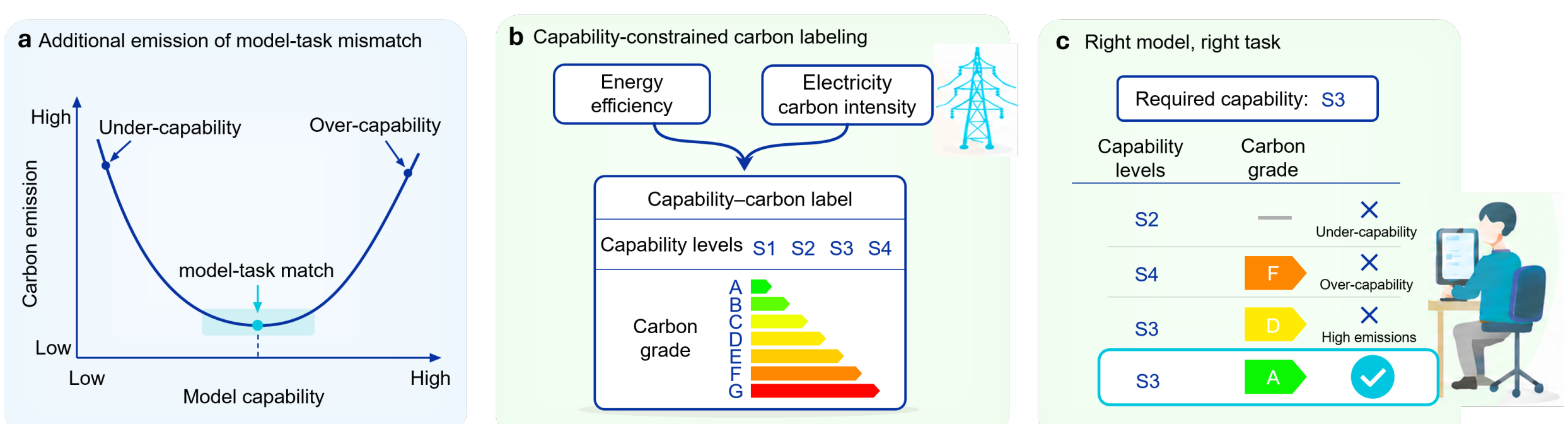

**Fig. 1 | Capability-constrained carbon labeling for right-model right-task matching. a**, Model-task mismatch increases carbon emissions per successful task through both under-capability and over-capability. **b**, Capability-constrained carbon labeling framework combines task-specific capability levels, energy efficiency and serving-region electricity carbon intensity. **c**, Users can select the lowest-carbon model at the required capability level, enabling right-model, right-task matching. For a task requiring capability level S3, under-capability models are excluded because they do not meet the required performance, whereas over-capability models may consume unnecessary computation. Among models that meet the

capability requirement, those with higher expected carbon emissions per successful task are disfavored, and the lowest-carbon eligible model is selected.

## Establishing Capability-constrained carbon labels

Capability-constrained carbon labeling should be built around three principles: service-based comparability, comprehensive energy accounting and dynamic carbon accounting. Together, these principles shift carbon labeling away from isolated measures of model efficiency towards the carbon emission of delivering a defined level of AI service.

First, service-based comparability requires models to be compared only after the task category and required capability level have been specified. Existing capability assessments do not establish the performance thresholds that distinguish capability levels within a task category. Therefore, capability levels should be defined within task categories using explicit performance or success criteria. Models that fail to meet the required capability level should not be considered equivalent alternatives, whereas models that meet or exceed that requirement can be compared on their environmental performance.

Second, comprehensive energy accounting requires evaluation of the total energy needed to deliver a successful task, rather than the energy associated with a single inference or final output. Failed attempts, repeated inference, retries and escalation can all increase the resources required before successful completion. Therefore, energy performance should be evaluated per successfully completed task at the required capability level, with task definitions, hardware configurations and inference settings reported consistently.

Third, dynamic carbon accounting requires carbon labels to reflect the electricity carbon intensity associated with the actual deployment of inference. The same task-level energy consumption can result in different emissions depending on where and when computation is performed. Task-execution energy should therefore be combined with the corresponding serving-region and, where available, time-specific electricity carbon intensity to estimate operational carbon emissions per successful task [8]. As a result, the lowest-energy option need not be the lowest-carbon option. Carbon labels should be interpreted as properties of model-task-deployment configurations rather than intrinsic properties of models, and should be updated as deployment conditions and electricity systems change.

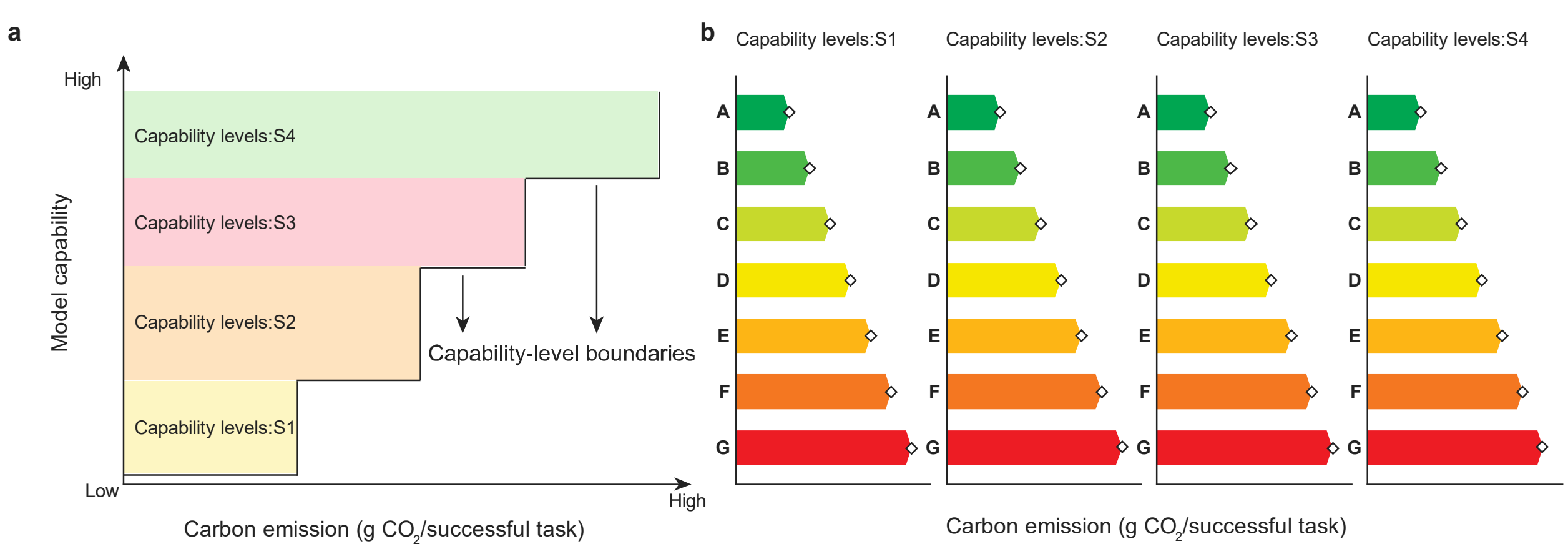

**Fig. 2 | Proposed construction of capability-constrained carbon labels for generative AI. a,** Task-specific performance requirements define four illustrative capability levels (S1-S4). Models are assessed against the corresponding service requirements to determine their eligibility for each level. **b**, Within each capability level, eligible models are classified into seven carbon grades (A-G) according to their expected operational carbon emissions per successfully completed task. Carbon emissions are determined by task-level energy consumption and serving-region electricity carbon intensity, with A indicating the lowest emissions and G the highest.

These principles can be translated into a practical labeling scheme, as illustrated in Fig. 2. Capability-constrained carbon labels should be developed separately for different task categories, such as writing, coding and mathematical reasoning, with capability levels defined by task-specific performance thresholds. In the illustrative scheme, S1-S4 denote increasing capability levels. Models are assessed against these thresholds to determine their eligibility. For a given required capability level, all models meeting or exceeding that level are compared according to their expected operational carbon emissions per successfully completed task. These emissions reflect both task-level energy requirements and the electricity carbon intensity associated with deployment. Drawing on the information structure of EU energy labels [7], we use grades A-G to represent increasing operational carbon emissions, with A indicating the lowest and G the highest (Fig. 2b). Capability levels therefore determine eligibility, whereas carbon grades distinguish among models capable of delivering an equivalent level of service. Models exceeding the minimum capability requirement remain eligible and are evaluated against the same service requirement. The capability-level boundaries and carbon-grade thresholds shown here are illustrative rather than prescriptive and would require calibration using standardized task-level performance, energy and carbon data.

## Label-guided model selection and emissions

Capability-constrained carbon labels aim to match the right model to the right task by jointly considering task-specific capability requirements and carbon emission per successful task. The objective is not to maximize model capability or minimize inference energy in isolation, but to minimize expected carbon emissions while maintaining the required level of service.

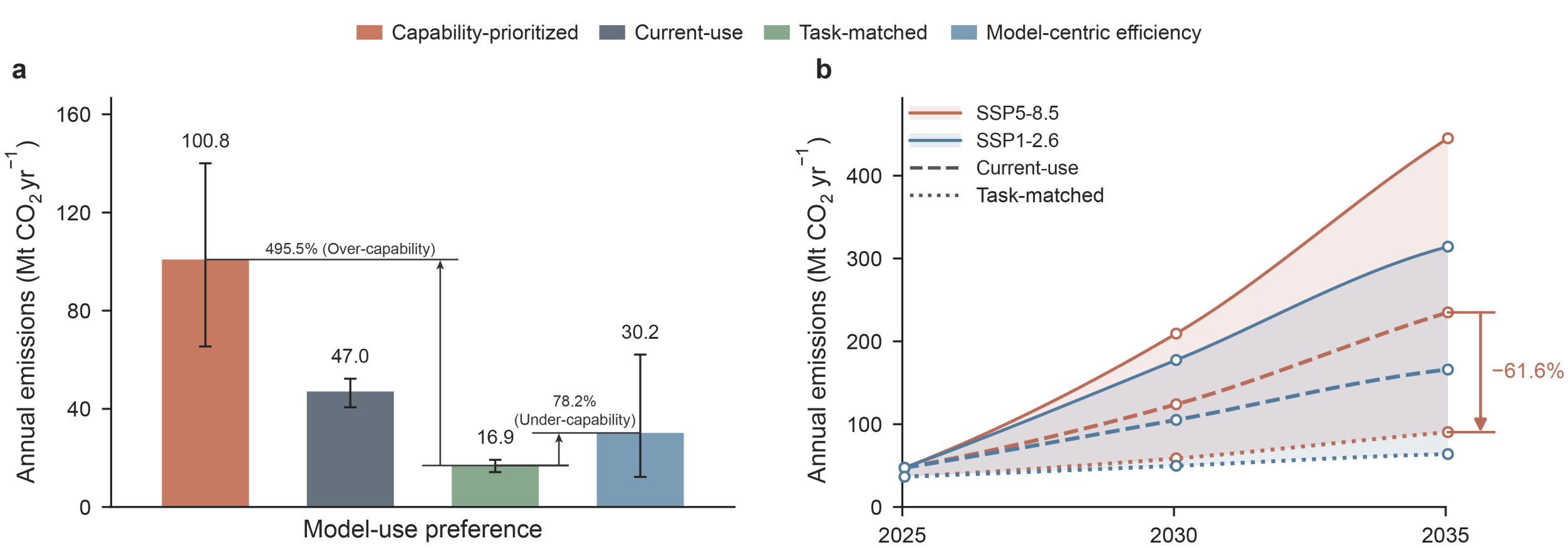

**Fig. 3 | Label-guided model selection can substantially reduce present and future emissions. a,** Estimated annual emissions in 2025 under four model-use scenarios: capability-prioritized, current-use, task-matched, and energy-efficiency-prioritized selection. The current-use distribution is derived from model-level token-use shares reported in [9], with models classified using published benchmark scores. The resulting usage shares across capability levels are assumed to approximate the distribution of service requirements. For energy-efficiency-prioritized selection, up to two attempts are allowed before escalation to a model meeting the required capability level, with energy consumption from unsuccessful attempts and

escalation included. Error bars reflect variations in estimated emissions arising from alternative capability-scoring systems [10]-[17]. **b,** Projected annual emissions from 2025 to 2035 under the same model-use scenarios for SSP1-2.6 and SSP5-8.5. The arrow indicates the relative emissions reduction achieved by task-matched selection compared with current use in 2035 under SSP5-8.5.

To illustrate the implications of different selection principles, we consider these choices through four model-use scenarios: a reconstructed current-use baseline, capability-prioritized selection, task-matched selection, and energy-efficiency-prioritized selection (Fig. 3). Published model-use shares provide a proxy for the baseline service-tier distribution [9]. Capability-prioritized selection favours the highest-performing model according to conventional benchmark rankings, whereas energy-efficiency-prioritized selection favours models with higher inference energy efficiency. Neither strategy explicitly imposes the capability level required for individual tasks. By contrast, task-matched selection identifies models that satisfy the required capability threshold before comparing their energy consumption per successfully completed task. We estimate counterfactual emissions under 2025 conditions and explore trajectories to 2035 using IEA-informed assumptions about the future growth of generative AI [18] and regional electricity carbon intensities under Shared Socioeconomic Pathway (SSP) 1-2.6 and SSP5-8.5 [19].

Two key implications emerge from this illustrative comparison. First, conventional model-level rankings can lead to substantial carbon penalties when the capability levels required by individual tasks are not explicitly considered. Under the assumed 2025 conditions, capability-prioritized selection results in 100.8 Mt $CO_2$, 495.5% higher than task-matched selection (16.9 Mt $CO_2$). Similarly, energy-efficiency-prioritized selection results in 30.2 Mt $CO_2$, 78.2% higher than task-matched selection, illustrating how model-level efficiency rankings may fail to capture the additional energy associated with unsuccessful attempts, retries and escalation (Fig. 3a). Second, capability-constrained model selection offers substantial opportunities for emissions reduction. Task-matched selection reduces estimated operational emissions by 64.0% relative to current use, from 47.0 to 16.9 Mt $CO_2$ (Fig. 3a). This advantage persists in illustrative pathways to 2035, with task-matched selection yielding 61.6% lower emissions than the current-use baseline under SSP5-8.5 (Fig. 3b).

Although the magnitude of these estimates depends on assumed task requirements, capability assessments and available inference-energy data, the comparisons highlight a broader conclusion: neither model capability nor inference energy efficiency alone determines the carbon emission of delivering a successful AI service. Instead, capability should serve as a task-specific eligibility constraint, with expected carbon emissions per successfully completed task guiding selection among models that reliably meet the required quality threshold. This shifts the basis of carbon labeling from evaluating models in isolation to assessing the carbon cost of successfully delivered AI services.

## Conclusion and outlook

Decarbonizing generative AI is not only a supply-side challenge of cleaner electricity and more efficient computing; it also requires addressing demand-side inefficiencies arising from mismatches between model capabilities and task requirements. Addressing this challenge requires moving beyond model-centric assessments towards evaluating AI services delivered at defined quality standards. Capability-constrained carbon labeling places task-specific service requirements at the centre of this assessment: model capability determines eligibility, while expected carbon emissions per successfully completed task provide a common basis for comparing eligible models. By linking service quality to carbon emission, this approach enables more informed matching of the right model to the right task.

Translating this framework into practice requires progress in three areas.

First, defining task-specific capability levels requires a clear and consistent understanding of what constitutes successful service. Unlike tasks with objectively verifiable outcomes, open-ended tasks such as writing and complex reasoning may require multidimensional criteria covering quality, reliability and safety. These criteria should be developed collaboratively by domain experts, end users and independent evaluators, rather than determined solely by model providers. Shared evaluation protocols should establish explicit performance thresholds for different capability levels, while retaining reference models and overlapping tasks to maintain comparability as models evolve.

Second, energy measurements should capture the full cost of successful task completion rather than isolated inferences. This requires consistent system boundaries that account for failures, retries and escalation where applicable, alongside transparent reporting of model, hardware and inference conditions. Independent verification is particularly important for closed-weight models.

Third, credible carbon accounting must translate task-level energy consumption into emissions that reflect the carbon intensity of electricity supplying inference, which varies across locations and over time. Carbon labels should therefore be deployment-specific, supported by transparent accounting methods and periodically updated as electricity systems evolve. Broader life-cycle impacts should also be incorporated as the evidence and reporting required for their assessment improve.

Together, these advances could establish a common basis for comparing the carbon cost of AI services delivered at equivalent quality standards. Although lower emissions per successful task may not translate into lower aggregate emissions if efficiency gains stimulate additional demand, capability-constrained carbon labeling offers a demand-side complement to cleaner electricity and more efficient computing. Its broader contribution lies in enabling carbon-informed model–task matching, ensuring that service requirements, rather than model capability or inference efficiency alone, guide the allocation of computational resources.

## References

[1] Alissa, H. *et al*. Using life cycle assessment to drive innovation for sustainable cool clouds. *Nature* **641**, 331–338 (2025).

[2] Colangelo, P. *et al*. AI data centres as grid-interactive assets. *Nat. Energy* **11**, 254–261 (2026).

[3] Xiao, T. *et al*. Environmental impact and net-zero pathways for sustainable artificial intelligence servers in the USA. *Nat. Sustain*. **8**, 1541–1553 (2025).

[4] Pham, D. *et al*. AgentStop: terminating local AI agents early to save energy in consumer devices. In *Proc. ACM Conference on AI and Agentic Systems (CAIS '26)* (2026).

[5] Chadli, K. *et al*. PEARL: performance and energy aware routing for LLMs. Future Gener. *Comput. Syst*. ***176***, 108218 (2026).

[6] Liu, J. *et al*. Is your code generated by ChatGPT really correct? Rigorous evaluation of large language models for code generation. *Adv. Neural Inf. Process. Syst.* **36** (2023).

[7] European Commission. Commission Delegated Regulation (EU) 2019/2016 of 11 March 2019 with regard to energy labelling of refrigerating appliances. *Off. J. Eur. Union* **L315**, 102–133 (2019).

[8] Google Cloud. *Use regions that consume low-carbon energy*. https://docs.cloud.google.com/architecture/framework/sustainability/low-carbon-regions (2026).

[9] OpenRouter. AI model rankings. https://openrouter.ai/rankings (2026).

[10] Epoch AI. *Epoch Capabilities Index*. https://epoch.ai/eci (2026).

[11] Ho, A. *et al*. A Rosetta Stone for AI benchmarks. Preprint at https://arxiv.org/abs/2512.00193 (2025).

[12] Rein, D. *et al*. GPQA: a graduate-level Google-proof Q&A benchmark. In *Proc. First Conference on Language Modeling* (2024).

[13] Epoch AI. GPQA Diamond. https://epoch.ai/benchmarks/gpqa-diamond (2026).

[14] Center for AI Safety, Scale AI & HLE Contributors Consortium. A benchmark of expert-level academic questions to assess AI capabilities. *Nature* **649**, 1139–1146 (2026).

[15] Pyatkin, V. *et al*. Generalizing verifiable instruction following. In *Advances in Neural Information Processing Systems* **38** (2025).

[16] Artificial Analysis. *Artificial Analysis Intelligence Benchmarking Methodology*. https://artificialanalysis.ai/methodology/intelligence-benchmarking (2026).

[17] Chiang, W.-L. *et al*. Chatbot Arena: an open platform for evaluating LLMs by human preference. *Proc. Mach. Learn. Res.* **235**, 8359–8388 (2024).

[18] International Energy Agency. *Energy and AI*. https://www.iea.org/reports/energy-and-ai (2025).

[19] O'Neill, B. C. *et al*. The Scenario Model Intercomparison Project (ScenarioMIP) for CMIP6. *Geosci. Model Dev.* **9**, 3461–3482 (2016).